\documentclass[a4paper,11pt,leqno]{article}

\usepackage{graphicx}
\usepackage{amsmath}
\usepackage{amssymb}
\usepackage{amsthm}
\usepackage{fullpage}
\usepackage{bm}
\usepackage{xcolor}
\usepackage{url}
\usepackage{hyperref}
\usepackage{setspace}
\usepackage{chngcntr}

\counterwithin{figure}{section}
\counterwithin{table}{section}
\numberwithin{equation}{section}

\hypersetup{ 
  pdfnewwindow=true, 
  colorlinks=true, 
  linkcolor=red,
  citecolor=brown, 
  filecolor=magenta, 
  urlcolor=cyan}

\numberwithin{theorem}{section}
\numberwithin{corollary}{section}
\numberwithin{proposition}{section}

\begin{document}
\title{Predicting the Next Maxima Incidents of the \\Seasonally Forced SEIR Epidemic Model}
\author{Charalampos A. Chrysanthakopoulos\footnote{Physics Department, University of Athens, Athens, Greece 
 (\href{mailto:cchrysanth@phys.uoa.gr}{cchrysanth@phys.uoa.gr})}}
\maketitle
\thispagestyle{empty}

\begin{abstract}
This paper aims at predicting the next maxima values of the state variables of the seasonal SEIR epidemic model and their in-between time intervals. Lorenz's method of analogues is applied on the attractor formed by the maxima of the corresponding state variables. It is found that both quantities are characterized by a high degree of predictability in the case of the chaotic regime of the parameter space.
\end{abstract}


\section{Introduction}
\label{sec:intro}

In the past decades, the focus of many studies has been to classify and understand the dynamics of the maxima - or 'Peak-to-Peak Dynamics' (PPD) \cite{Cand00} - of various dynamical systems. In biological systems and specifically in disease spread models, the maxima incidents, i.e. incidents where at least one state variable has maximum value, are of primary concern. A classification tool widely used in this field of research is the $(x_{n},x_{n+1})$ plot, where $x_n$ are the state variables' maxima values of the corresponding dynamical system. These plots are called Peak-to-Peak-Plots or PPPs. In the event that the points are dense and form simple lines, the plots are often called \emph{filiform}, while when the points form a widespread figure the plots are \emph{non-filiform}, as illustrated in figure \ref{fig:d_attractor}. In regards to PPD, great interest has been drawn in the development of predictive methods. The R{\"o}ssler hyperchaotic system, the EEG of an individual at rest and the seasonal SEIR epidemic model, are examples of systems that exhibit non-filiform PPP's. The prediction of the maxima values of the R{\"ossler hyperchaotic system has been investigated with nonlinear methods of prediction and it has been found that the maxima are characterized by a high degree of predictability \cite{Kug08}. This means that the prediction of the maxima values of systems that exhibit non-filiform PPPs can be tackled by standard nonlinear predictive methods. The prediction of the annual maxima values of the population of infectives of the SEIR model has already been discussed \cite{Tidd93}. In the present study, the focus is extended to all state variables of the SEIR model as well as the time intervals between successive maxima values for each variable. 

\section{Setup}
\label{sec:Setup}

Let us consider a fixed size population of susceptible, exposed, infective and recovered individuals. Let $S$, $E$, $I$ and $R$ denote the fraction of individuals for each corresponding class, so that $S+E+I+R=1=constant$. 
We assume that new susceptible individuals are introduced at a constant birth rate $\mu$ equal to the death rate, so that the population under study is of fixed size. We also consider a rate of disease transmission $\beta$.
Moreover, we accept that exposed individuals become infective at a rate $\alpha$ and recover at a rate $\gamma$, thus becoming permanently immune.
These assumptions result in the development of the SEIR epidemic model

\begin{eqnarray}
 \label{eq:SEIR1}\frac{dS}{dt} & = & \mu-\mu S-\beta SI \\
 \frac{dE}{dt} & = & \beta SI-(\mu + \alpha)E \\
 \label{eq:SEIR3}\frac{dI}{dt} & = & \alpha E-(\mu + \gamma)I 
\end{eqnarray}

The fraction of recovered individuals can easily be estimated from $R=1-S-E-I$. Dietz \cite{D76} proposed parameter $\beta$ to be time-dependant in order to encompass seasonal variations. Following Kuznetsov \emph{et al.} \cite{Kuz94}, we introduce a sinusoidal seasonal forcing $\beta=\beta_0 [1+\delta \cdot cos(2 \pi t)]$, where $\delta$ is called 'degree of seasonality'. Quantity $B=\beta_0 \cdot \delta$ is crucial for the dynamics of the system in the sense that it plays the role of the external forcing amplitude. Specifically, the greater the amplitude of the forcing, the greater the pertubation to the system, i.e. for $B < B_{threshold}$ the state variables' maxima inhabit on stable-period cycles, while for $B > B_{threshold}$ a transition takes place from regular stable dynamics to chaotic dynamics via a period-doubling route to chaos \cite{Kuz94, Tidd93}. Naturally, $B_{threshold}$ depends on the values of the other parameters of the system, which are constrained from historical data that the sytem is built to simulate.

Other $\beta(t)$ functions have been proposed \cite{Kot88}, however the general behavior of the system is asymptotically chaotic regardless of the exact $\beta(t)$ expression, as long as it is sinusoidal.

\section{Methodology and Implementation}
\label{sec:Meth}

We choose the parameter values $\alpha=35.842~\frac{1}{year}, \beta_0=1884.95~\frac{1}{year}, \gamma=100~\frac{1}{year}, \delta=0.255$ and $\mu=0.02~\frac{1}{year}$ as in \cite{Kuz94}. We then integrate all state variables numerically with the lsoda method and integration step $\tau= \mathcal{O}(10^{-3})$. Furthermore, throughout this study we use initial conditions $S_0=0.05, E_0=0$ and $I_0=0.016$.

\begin{figure}[h!]
   \centering   
   \includegraphics[height=0.6\linewidth, width=0.8\linewidth]{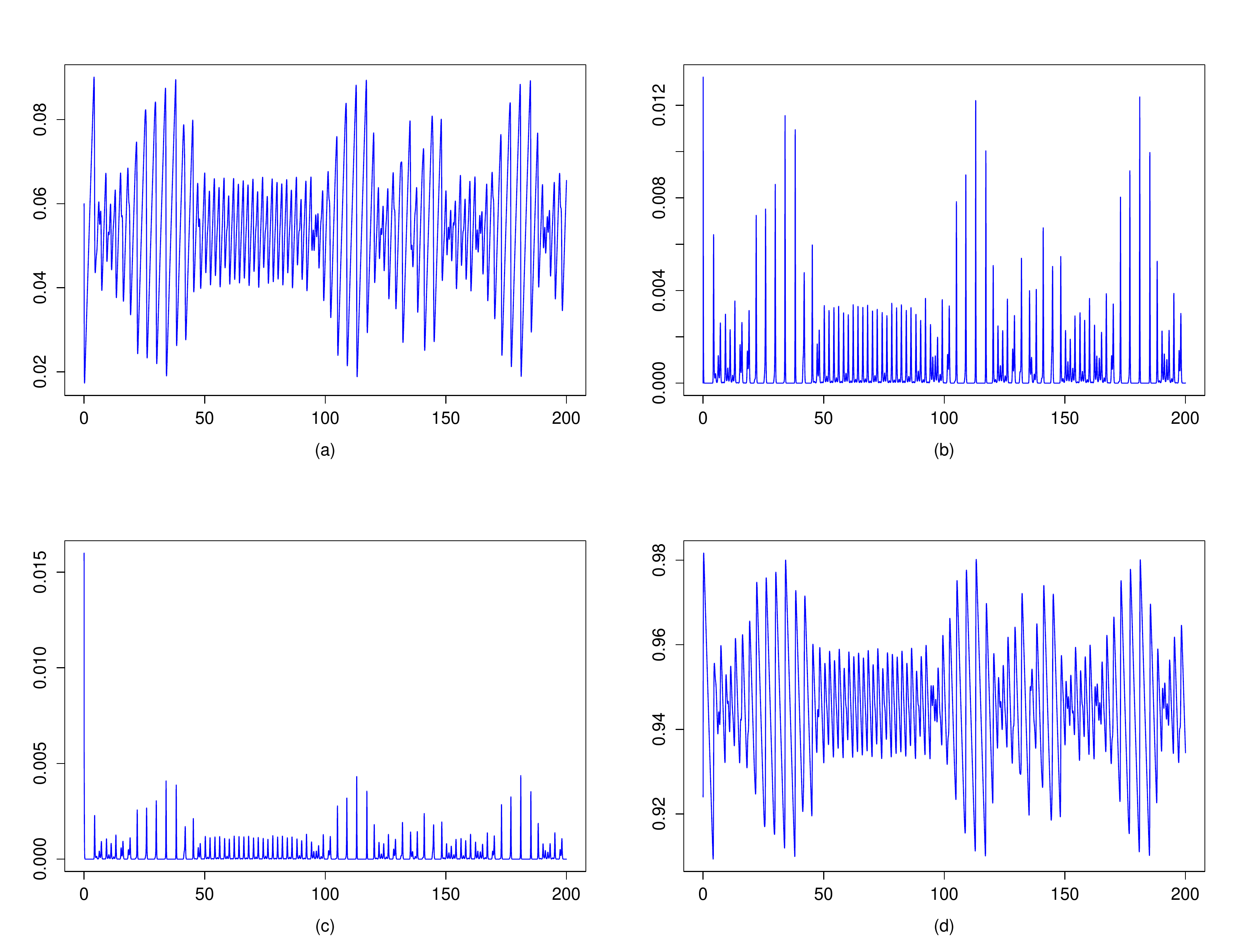}
   \caption{\label{fig:S_I_vs_t} Variables : (a) $S$, (b) $E$, (c) $I$ and (d) $R$ for $2 \cdot 10^4$ time steps.}
\end{figure}

The maxima values $S_{max}, E_{max}, I_{max}$ and $R_{max}$ are then extracted from each state variable $S, E, I$ and $R$ and the time intervals $\hat\tau_i=t_{i+1}-t_{i}$ between consecutive maxima incidents are calculated, where $t_i$ is the time of occurrence of the $i_{th}$ maximum. $S_{max,n+1}$ with respect to $S_{max,n}$ that is depicted in figure~\ref{fig:d_attractor}, is non-filiform, which makes it difficult to predict the next value of $S_{max}$ using methods that apply to filiform PPD dynamics as in \cite{Cand00} [also $(E_{max,n},E_{max,n+1})$, $(I_{max,n},I_{max,n+1})$ and $(E_{max,n},E_{max,n+1})$ plots are non-filiform]. This issue is addressed applying the phase space embedding technique which is based on Floris Takens' delay embedding theorem \cite{Tak81,Pack80}. Specifically, we embedd an initially scalar time series $\{x_n\}_{n=1}^{N}$ in a $m$ dimensional phase space constructing thus $\ell=N-(m-1) \cdot d$ vectors $\textbf{x}_j$, so that a vectorial time series $\{ \textbf{x}_n \}_{n=1}^{\ell}$ is produced. Parameter $d$ is called 'delay time' and helps formulate each embedded vector like $\textbf{x}_n=(x_{n}, x_{n+d}, ..., x_{n+(m-1) \cdot d})$, so that every vector consists of $m$ components. 

\begin{figure}[h!]
   \centering   
   \includegraphics[height=0.4\linewidth, width=0.8\linewidth]{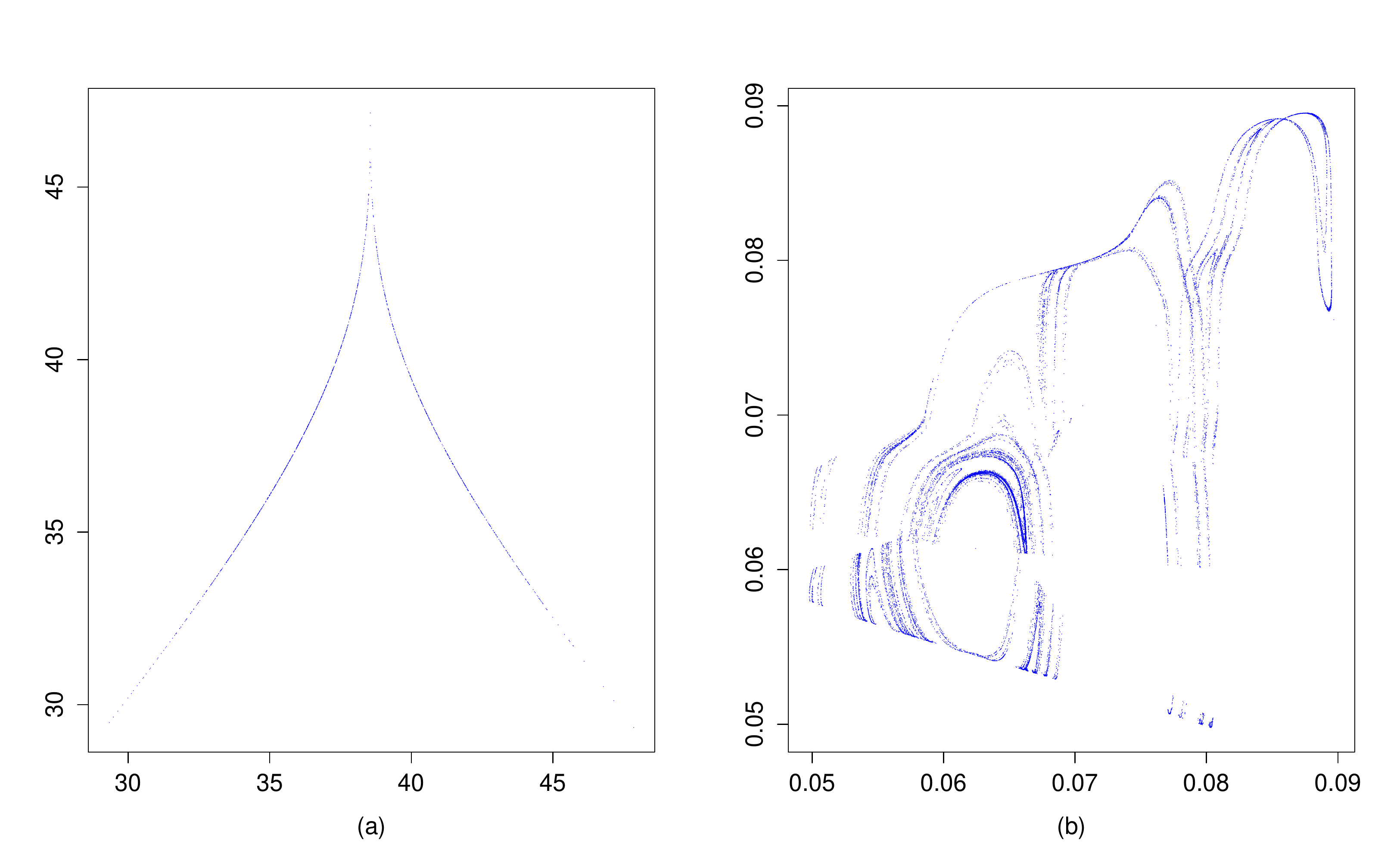}
   \caption{\label{fig:d_attractor} (a) Filiform PPP: $1333$ points of $(z_{max,n}, z_{max,n+1})$, where $z_{max,n}$ are the maxima of $z$ component of the Lorenz system for $(\sigma, \rho, \beta)=(10, 28, -8/3)$ and initial conditions $(x_0, y_0, z_0)=(0.1, -0.3, 1.7)$. (b) Non-filiform PPP: $18001$ points of $(S_{max,n}, S_{max,n+1})$, where $S_{max}$ are the maxima of variable $S$ of seasonal SEIR model.}
\end{figure}

The predictive scheme proposed has been used for the prediction of influenza spread from historical data \cite{Vib03}. It is a method based on an idea first proposed by Edward Lorenz \cite{L69}, namely 'Lorenz's method of analogues'. According to the method, the next value of an initially scalar time series $\{x\}_{n=1}^{N}$ is estimated
\begin{equation}
 x_{N+1}=\textbf{x}_{\ell+1}^{(m)}=\frac{1}{K} \cdot \sum_{j=1}^K \textbf{x}_{j+1}^{(m)}
\end{equation}
where $\textbf{x}_{\ell + 1}^{(m)}$ is the $m^{th}$ component of the future embedded vector $\textbf{x}_{\ell + 1}$. $\textbf{x}_{\ell}$ is the last embedded vector so that $\textbf{x}_{\ell}^{(m)}=x_N$ and $K$ is the number of the nearest $\textbf{x}_j$ neighbor vectors to $\textbf{x}_{\ell}$. In other words, future value $x_{N+1}$ should be close to the average value of the $m^{th}$ components of the $K$ neighbors' future vectors $ \{ \textbf{x}_{j+1} \}_{j=1}^{K} $.

We prefer this method because it outperforms higher polynomial methods \cite{Tidd93}. Also, we avoid using the 'sphere' variant of the method, namely choosing all neighbors within a sphere of radius $\rho$ from the last embedded vector, in order to avoid taking into account neighbors that may contaminate our prediction. The danger for this originates from the fact that the neighbor density in the phase space is not constant, so considering a constant radius for our model may introduce neighbors irrelevant to the local phase space dynamics.

We study both the maxima values $S_{max}, E_{max}, I_{max}, R_{max}$ and the first differences of the times of occurrence of the maxima values $\hat{\tau}_{i}$ for each variable (see figure \ref{fig:time_plots}).

\begin{figure}[h!]
   \centering   
   \includegraphics[height=0.4\linewidth, width=0.8\linewidth]{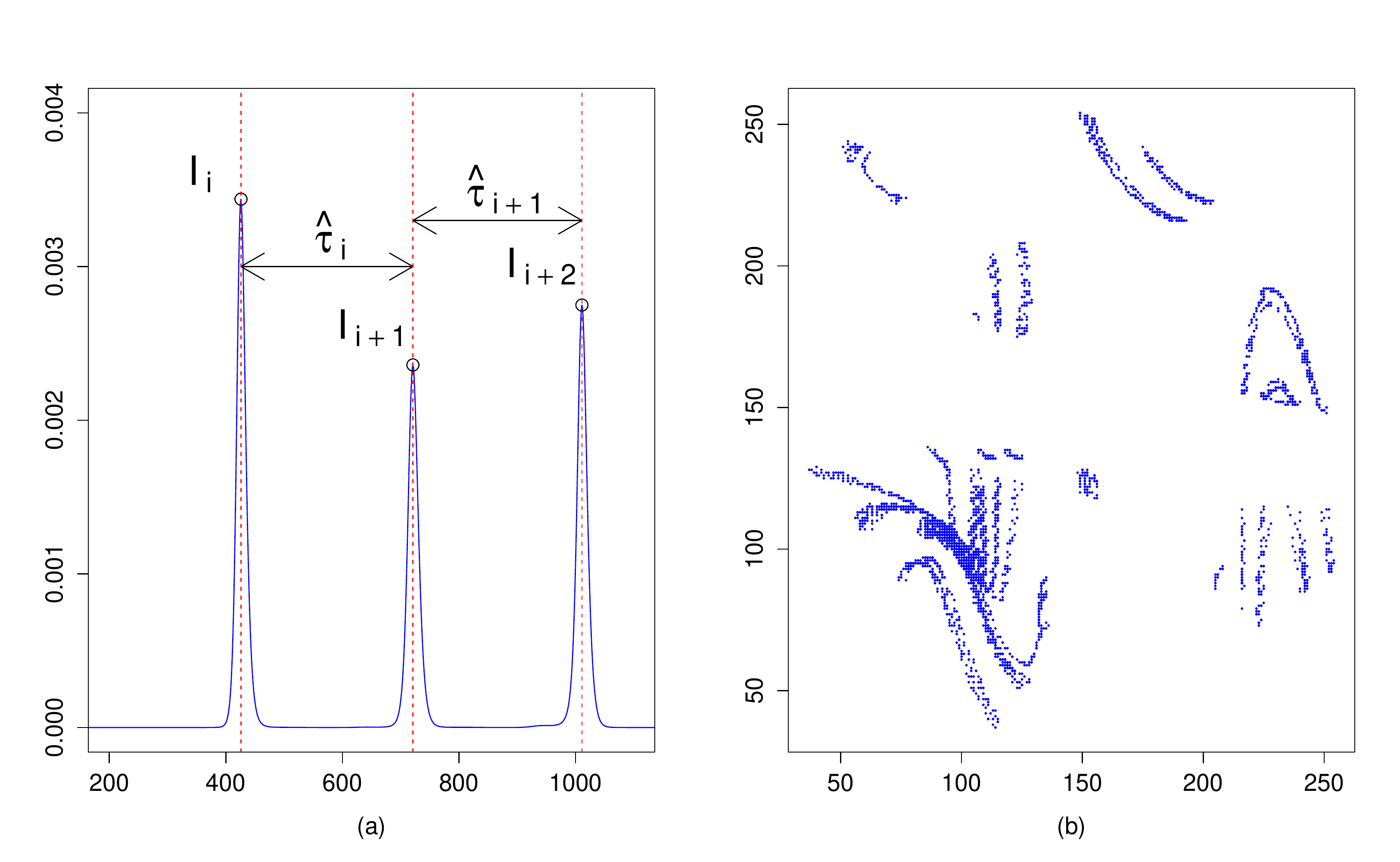}
   \caption{\label{fig:time_plots} (a) Three consecutive maxima incidents $I_{max}$ (black circles) of the population of infectives $I$ (blue line) and the intervals $\hat{\tau}_i$ between them, for $1100$ time steps. (b) 8300 points $(\hat{\tau}_i, \hat{\tau}_{i+1})$ of the corresponding maxima incidents $I_{max}$.}
\end{figure}

Finally, in order to assess the predictive power of the model $\textbf{M}( \cdot )$, the Nash - Sutcliffe model efficiency coefficient is used
\begin{equation}
 NS(x, x_{pred}; T)=1-\frac{\sum_{n=1}^T\left(x_n-x_{pred,n}\right)^2}{\sum_{n=1}^T\left(x_n-\overline{x}\right)^2}
\end{equation}
where $T$ is the test period length, $x_n$ are the real values, $x_{pred,n}$ are the predicted values and $\overline{x}$ is the average of the real values over the test period. If $NS$ is close to $1$, the prediction is successful, while for $NS$ close to $0$ it is equally accurate to the mean of the real data. If $NS<0$, the prediction is unsuccessful. 

Our predictive method depends on the length $M$ of the 'training' time series $\{x_{n}\}_{n=1}^{M}$, that is the time-series that is stored as historical data. It is important to choose $M$ big enough to include all behaviors of the time-series we wish to predict.

\section{Results}
\label{sec:Res}

We calculated the Nash-Sutcliffe coefficient over a test period of length $T=80$. For each test $719$ to $799$ values were used as historical data for maxima values and time intervals $\hat\tau_i$, from first to last predicted point correspondingly. For the embedding we choose $d=1$ and various values for the embedding dimension $m$. Also the number of nearest neighbors $NN$ varies as presented in tables $\ref{table:tab1}$ and $\ref{table:tab2}$.  

We did not use the optimal model $\textbf{M}_{opt}(\cdot)$ that produces the best prediction over the test period $x_{test}$, i.e. the maximum Nash-Sutcliffe coefficient $NS_{max}$, to carry out an out of sample forecasting over $x_{out-of-sample}$, since the optimal model  for $x_{test}$ does not always produce the best prediction for $x_{out-of-sample}$. Expressly, if $\textbf{M}_{opt} (x_{test}) \implies NS_{max} $ and $ \textbf{M}'_{opt} (x_{out-of-sample}) \implies Best~Out~of~Sample~Forecasting$, then in most cases $\textbf{M}_{opt}( \cdot ) \neq \textbf{M}'_{opt}( \cdot ) $, even though $\textbf{M}'_{opt}( \cdot )$ can produce results in the vicinity of the ones that $\textbf{M}_{opt}( \cdot )$ does. 

\begin{table}[ht]
\centering
\begin{tabular}{l*{11}{c}r}
  	                              & & $m=2$ &  &\vline &  & $m=3$ &  &\vline &  & $m=4$ & &   \\
  \hline 
  $NN$				    & $1$ & $2$ & $3$ &\vline & $1$ & $2$ & $3$ & \vline& $1$ & $2$ & $3$ \\
  \hline 
  $S_{max}$     & 0.884 & 0.911 & 0.881 & \vline & 0.770 & 0.767 & 0.705 & \vline & 0.631 & 0.571 & 0.528 \\ 
  $E_{max}$	    & 0.822 & 0.869 & 0.799 & \vline & 0.879 & 0.945 & 0.917 & \vline & 0.980 & 0.971 & 0.983 \\ 
  $I_{max}$	    & 0.795 & 0.843 & 0.798 & \vline & 0.913 & 0.926 & 0.915 & \vline & 0.961 & 0.970 & 0.983 \\ 
  $R_{max}$	    & 0.938 & 0.955 & 0.927 & \vline & 0.970 & 0.900 & 0.861 & \vline & 0.914 & 0.794 & 0.714 \\   
\end{tabular}
\caption{Values of Nash-Sutcliffe coefficient $NS$ for the state variables' maxima values.}
\label{table:tab1}
\end{table}

\begin{table}[ht]
\centering
\begin{tabular}{l*{11}{c}r}
  	                              & & $m=2$ &  &\vline &  & $m=3$ &  &\vline &  & $m=4$ & &   \\
  \hline 
  $NN$				    & $1$ & $2$ & $3$ &\vline & $1$ & $2$ & $3$ & \vline& $1$ & $2$ & $3$ \\
  \hline 
  $\hat\tau_{S_{max}}$     & 0.667 & 0.634 & 0.636 & \vline & 0.679 & 0.613 & 0.644 & \vline & 0.610 & 0.550 & 0.484 \\
  $\hat\tau_{E_{max}}$	   & 0.989 & 0.995 & 0.961 & \vline &0.990 & 0.996 & 0.960 & \vline & 0.898 & 0.951 & 0.917 \\
  $\hat\tau_{I_{max}}$	   & 0.894 & 0.911 & 0.888 & \vline &0.892 & 0.909 & 0.919 & \vline & 0.908 & 0.952 & 0.919 \\
  $\hat\tau_{R_{max}}$	   & 0.806 & 0.813 & 0.837 & \vline & 0.876 & 0.864 & 0.833 & \vline & 0.835 & 0.826 & 0.767 \\   
\end{tabular}
\caption{Values of Nash-Sutcliffe coefficient $NS$ for the time intervals $\hat\tau$ between the state variables' maxima values.}
\label{table:tab2}
\end{table}


\section{Conclusion}
\label{sec:Conc}

It is established that the prediction of the maxima values for all populations of the seasonally forced SEIR epidemic model is possible, result that coincides with other studies \cite{Tidd93}. The time intervals in-between successive maxima incidents $\hat\tau_i$ are also characterized by high predictability, therefore the time of occurrence of each maximum value can be determined, since $t_{i+1}=\hat\tau_i + t_i$. Future research can focus on the efficacy of the method on real world data and consider different parameter values.

\end{document}